\documentclass[prd, twocolumn, nofootinbib, showpacs, superscriptaddress]{revtex4-1}
\usepackage[utf8]{inputenc}
\usepackage{graphicx}
\usepackage{color}
\usepackage{amsmath}
\usepackage{hyperref}
\usepackage{tabularx}
\usepackage{amsmath,amssymb}
\usepackage{float}
\usepackage{subcaption}

\usepackage[normalem]{ulem}
\newcommand{\be}{\begin{equation}}
\newcommand{\ee}{\end{equation}}
\newcommand{\bea}{\begin{eqnarray}}
\newcommand{\eea}{\end{eqnarray}}

\newcommand{\gapp}{\mathrel{\raise.3ex\hbox{$>$}\mkern-14mu \lower0.6ex\hbox{$\sim$}}}
\newcommand{\lapp}{\mathrel{\raise.3ex\hbox{$<$}\mkern-14mu \lower0.6ex\hbox{$\sim$}}}
\def\bbox{{\,\lower0.9pt\vbox{\hrule \hbox{\vrule height 0.2 cm
\hskip 0.2 cm \vrule  height 0.2 cm}\hrule}\,}}

\usepackage{color}
\usepackage{ulem}
\usepackage[usenames,dvipsnames]{xcolor}

\begin{document}
\title{Gravitational wave echoes from black holes in massive gravity}

%\author{Vitor Cardoso}
%\email[Email: ]{vitor.cardoso@ist.utl.pt}
%\address{CENTRA, Departamento de F\'isica, Instituto Superior T\'ecnico,
%Universidade de Lisboa, Avenida Rovisco Pais 1, 1049 Lisboa, Portugal}
\author{Ruifeng Dong}
\email[Email: ]{ruifengd@buffalo.edu}
%\address{HEPCOS, Department of Physics, SUNY at Buffalo, Buffalo, NY 14260-1500}
\author{Dejan Stojkovic}
\email[Email: ]{ds77@buffalo.edu}
\address{HEPCOS, Department of Physics, SUNY at Buffalo, Buffalo, NY 14260-1500}

 %%%%%%%%%%%%%%%%%%%%%%%%%%%%%%%%%%%%%%%%%%%%%%%%%%%%%%%

\begin{abstract}
Gravitational waves are rapidly becoming a very reliable tool for testing alternative theories of gravity. In particular, features in the gravitational wave emission during black hole ringdown phase provide a direct probe of the spacetime outside the black hole. In this article, we consider the Schwarzschild-de Sitter black hole solution in ghost-free massive gravity. These black holes generically have scalar hair. We found that a simple coupling between gravitational perturbations and this scalar hair can change the quasinormal ringing of the black hole, and in particular, produce echoes in the emitted gravitational waves. This finding suggests a possible method for testing massive gravity using gravitational wave observations.
\end{abstract}

%%%%%%%%%%%%%%%%%%%%%%%%%%%%%%%%%%%%%%%%%%%%%%%%%%

\pacs{}
\maketitle

\section{Introduction}

In 1967, Ames and Thorne \cite{AmesThorne} found that the photosphere of a black hole can maintain unstable circular photon orbits, which can release light even after the formation of event horizon in a gravitational collapse. This light-ring behavior was later found to be also responsible for the long-wave train by Press \cite{Press:1971wr}. In 1991, Chandrasekhar and Ferrari \cite{Chandrasekhar:1991fi} calculated the quasinormal modes for relativstic stars with $r_0<3M$. For the dynamics of a typical curvature perturbation, they found a minimum in the potential between the gravitational barrier at the photosphere and the centrifugal barrier closer to the center of the star. A number of quasistationary quasinormal modes were found. In 2000, Ferrari and Kokkotas \cite{Ferrari:2000sr} extended the calculation to the time domain and found the appearance of a series of later-time signals after the quasinormal ringdown in the strain time series. These are called gravitational-wave echoes.

The appearance of gravitational-wave echoes can be understood intuitively as follows. The gravitational perturbation can be trapped between the two peaks in the effective potential. Everytime the perturbation bounces at the right boundary of the peak and partially penetrates through, we expect to receive an echo by the detector at a later time determined roughly by the proper distance between the two peaks.

The principal quasinormal mode was clearly observed in the gravitational waves produced during the formation of a final black hole in the binary-merger events observed in the first run of aLIGO \cite{TheLIGOScientific:2016pea}. However, Cardoso et. al. \cite{Cardoso:2017cqb} cautioned that this ringdown mode is not a defining signature of the event horizon, but rather of the photosphere, which very compact objects can also possess \cite{Chandrasekhar:1991fi}. A list of extremely compact stars with clean photospheres were identified, which would produce echoes after the typical ringdown indistinguishable from that of black holes. Importantly, quantum structures near the gravitational radius, e.g. gravistars, fuzzballs and firewalls, are among this class of extremely compact stars \cite{Saraswat:2019npa}. Observation of echoes can, therefore, give hints on the black hole information paradox. 

Moreover, echoes can also be a testbed for quantum gravity. Quantization of the area of event horizon results in selectivity in the wave frequency absorbable by the black hole. This can also modify the picture of quasinormal ringdown, and in particular, produce echoes at certain frequencies \cite{Agullo:2020hxe}. 

A tentative framework of testing the existence of echoes has been put forward \cite{Saraswat:2019npa,Abedi:2016hgu}. Applications to the three events in the first run of aLIGO showed possible but inconclusive echo signals \cite{Abedi:2016hgu,Nielsen:2018lkf}. Further improvements in both observation and techniques of statistical testing will provide more exciting results, whether positive or negative. For a comprehensive overview of this subject, please refer to \cite{Cardoso:2019rvt}.

To the best of our knowledge, gravitational wave echoes are currently thought to be  possible only in wormholes \cite{Cardoso:2017cqb}, quantum structure near the event horizon \cite{Saraswat:2019npa, Cardoso:2019rvt}, or quantum gravity effects \cite{Agullo:2020hxe}. Here we propose a new possibility. In modified gravity theories, black holes might have different dynamical properties from black holes in general relativity. This difference might be reflected in gravitational waves emitted when a perturbed black hole relaxes down to its equilibrium state. 

It was recently found in \cite{Dong:2015qpa} that a black hole in massive gravity theory has drastically different greybody factors from black holes in general relativity. In particular, scalar perturbations to the black hole can couple not only to the black hole metric but also to non-trivial background fields, as we will show explicitly in Eqs. (\ref{Ai}, \ref{A0}, \ref{g(r)}). This adds to the greybody factor resonance at discrete frequencies, and enhances Hawking radiation at these frequencies. Physically, the greybody factor is the transmission probability of a mode in the BH Hawking radiation. Just like the BH ringdown signal, it is determined by the effective potential for the mode as well as the boundary conditions at the event horizon and at infinity. For the greybody factor, the mode is a mixture of incoming and outgoing waves at the event horizon and a purely outgoing wave at infinity. For the BH ringdown, the gravitational mode is purely incoming at the event horizon and a mixture of incoming and outgoing components at infinity.

In \cite{Dong:2015qpa}, it was found that the coupling between a scalar test field and the background field results in a double-peak effective potential. This then causes a resonant transmission at certain frequencies determined roughly by the proper distance between the
two peaks. Here as discussed above in the second paragraph, a similar coupling between the gravitational perturbation and the background field can potentially produce a double-peak effective potential for the gravitational mode and therefore cause echoes to appear in the BH ringdown signal.

Motivated by the above argument, we now look into the gravitational wave properties of such black holes when they are relaxing to their equilibrium state.

The paper is organized as follows. In section \ref{sec:mg}, we review the ghost-free massive gravity theory and its black hole solutions. Then in section \ref{sec:pert}, we will discuss gravitational perturbations of this black hole. The effects of background fields on the perturbations will be discussed afterwards in section \ref{sec:coup}. We will then compute the gravitational wave signals to be observed in asymptotic regions. In section \ref{sec:ana}, we will show an analytic example to look into the quasinormal ringing properties. Then in section \ref{sec:num}, we will discuss numerical calculations for the general case. The conclusion of this paper will follow afterwards in section \ref{sec:con}.

In this paper, we will use the natural units, i.e. $\hbar$=1, c=1, G=1.

\section{Black holes in massive gravity theory}\label{sec:mg}

We will consider the ghost-free massive gravity theory formulated in 2010 \cite{deRham:2010kj}. The Lagrangian takes the form
\be
\mathcal{L}= \frac{\sqrt{-g}}2\left( R+{m^2_g} \mathcal{U}(g_{\mu\nu},\phi^a)  \right),
\label{lagrangian}
\ee
where $m_g$ is graviton's mass. $g_{\mu\nu}$ is the metric tensor, $g$ is its determinant and $R$ is the corresponding Ricci scalar. $\phi^a$ are the St\"uckelberg fields and $\mathcal{U}$ contains their coupling with the metric. If we define a matrix $\mathcal{K}^{\mu}_{\nu}(g_{\alpha\beta},\phi^a)\equiv\delta^{\mu}_{\nu}-\sqrt{g^{\mu\alpha}\partial_{\alpha}\phi^a\partial_{\nu}\phi^b\eta_{ab}}$, then $\mathcal{U}$ takes the form
\be
\mathcal{U}(g,\phi^a)=\mathcal{U}_2 + \frac{\alpha-1}{3}\mathcal{U}_3 -\left(\frac{\eta}{2}+\frac{\alpha-1}{12}\right)\mathcal{U}_4,
\ee
where $\alpha$, $\eta$ are free parameters, and
\bea
\mathcal{U}_2&= &[\mathcal{K}]^2-[\mathcal{K}^2], \nonumber \\
\mathcal{U}_3&= &[\mathcal{K}]^3-3[\mathcal{K}][\mathcal{K}^2]+2[\mathcal{K}^3], \nonumber \\
\mathcal{U}_4&= &[\mathcal{K}]^4-6[\mathcal{K}^2][\mathcal{K}]^2+8[\mathcal{K}^3][\mathcal{K}]+3[\mathcal{K}^2]^2-6[\mathcal{K}^4].  \nonumber\\
\eea
Here the square bracket around a matrix denotes its trace.

The Stuckelberg fields are not fundamental fields but auxiliary fields to restore the certain symmetry in the theory. It was originally proposed to build models for massive photons, where Stuckelberg fields help restore the U(1) gauge invariance which is otherwise lost. It can be integrated out but the theory is equivalent but nonlinear as a result. See reference \cite{Ruegg:2003ps}. In massive gravity, Stuckelberg fields help keep the diffeomorphism invariance. We can get two equivalent descriptions of the theory, either you keep Stuckelberg fields and take all the couplings properly into account, or integrate it out but the theory is appropriately modified. We choose to work with the first option. For the latter one, please refer to \cite{deRham:2010kj} for the ADM formalism of the Lagrangian in Eq. (\ref{lagrangian}) without utilizing Stuckelberg fields.

In the special case $\eta=-\alpha^2/6$, this theory admits a static spherically symmetric solution, which was found in \cite{Berezhiani:2011mt}. The metric there takes the Schwarzschild-de-Sitter (SdS) form,
\be
ds^2=-f(r)dt^2+1/f(r){dr^2}+r^2d\Omega^2,
\label{ds2}
\ee
where $f(r)=(1-\frac{2M}r-\kappa^2m^2_g r^2)$. M is the mass of the black hole and $\kappa^2=\frac{2}{3\alpha}$. We see that the de Sitter (dS) horizon size depends on graviton's mass.

Along with the metric, we also have background fields, which are the vector and scalar modes of $\phi^a$, denoted as $A^{\mu}$ and $\pi$. Explicitly, $\phi^a=x^a-(m_g A^a + \partial^a\pi)/(M^2_{pl}m^2_g)$\cite{Berezhiani:2011mt}. In Schwarzschild coordinates, $A^{\mu}$ was found to be
\bea
A^i&=&0 , \label{Ai}\\
A^0&=& -\frac{M_{pl}m_g}{\kappa_0}g(r) , \label{A0}
\eea
and $\pi$ varies with both $r$ and $t$. Here the free dimensionless parameter $\kappa_0$ is an integration constant \cite{Berezhiani:2011mt}, and
\be
g(r)= \pm \int \frac{\sqrt{1-f(r)}}{f(r)} dr.
\label{g(r)}
\ee
Note that $g(r)$ diverges at both roots of $f(r)$, i.e. at both the black hole event horizon and cosmological horizon.

Note that the black hole solution in this theory is described by a Schwarzschild-de Sitter metric (\ref{ds2}) along with a form of the Stuckelberg fields in the Schwarzschild coordinates (\ref{Ai}, \ref{A0}, \ref{g(r)}). Due to the coupling between $g_{\mu\nu}$ and $\phi^a$ in the lagrangian in Eq.(\ref{lagrangian}), a perturbation in this theory is inevitably coupled both to the background Schwarzschild-de Sitter metric but also the background solution of $\phi^a$. This is the reason why $\phi^a$ can affect the quasinormal modes and ringdown signal of the black hole. 

Before moving forward, we want to stress that the scalar and vector modes in massive gravity are infinitely strongly coupled in the black hole solution above \cite{Berezhiani:2011mt}. We do not aim to perform a complete perturbation theory, but instead to search for interesting phenomenology based on some assumptions on the interaction between the helicity-2 part of perturbation and the background fields. This would be based on some assumptions as we shall discuss in the next section.

\section{Gravitational perturbations}\label{sec:pert}

In massive gravity theory, diffeomorphism invariance is lost and we therefore no longer have the gauge freedom when performing perturbations. While the current state-of-the-art in the field does not allow us to carry out the full perturbation theory in massive gravity, here we make the following assumptions. 

We treat massive gravity as a small correction to general relativity, in the sense that diffeomorphism invariance in the black hole solution is lost only at second order or higher orders. Then we can still use the gauge freedom in first-order perturbations, just as in general relativity. Afterwards, we will consider possible couplings between the perturbations and background Stuckelberg fields.

Note that the theory of massive gravity is not diffeomorphism invariant. We made the assumption above so as to be able to reveal possible effects on the gravitational waves from graviton mass. Further work is needed to justify or abandon this assumption.

We firstly review the treatment of gravitational perturbations around a SdS black hole in general relativity. A linear perturbation is added to the black hole metric, i.e.
\be
g_{\mu\nu}(x^{\lambda}) = g^0_{\mu\nu}(r) + h_{\mu\nu}(x^{\lambda}),
\ee
where $g^0_{\mu\nu}$ is the background metric and $h_{\mu\nu}$ is the perturbation.

The equations of motion of $h_{\mu\nu}$ can be derived in principle by putting the above perturbed metric into the vacuum Einstein equation. The difficulty arises due to the dependence and coupling between different components of $h_{\mu\nu}$. The Regge-Wheeler formalism deals with this issue. The perturbation can be separated into independent modes, each of which has a definite parity and angular momentum, due to the underlying reflection and rotational symmetries of the background. Moreover, a proper gauge can be freely chosen to reduce the dependence of different components of a mode.

In the Regge-Wheeler gauge, $h_{\mu\nu}$ takes the form \cite{Regge:1957td,Zerilli:1971wd}
\be
h^{odd}_{\mu\nu} = 
\begin{pmatrix}
0 & 0 & 0 & h_0(t,r) \\
0 & 0 & 0 & h_1(t,r) \\
0 & 0 & 0 & 0      \\
h_0(t,r) & h_1(t,r) & 0 & h_0(t,r)
\end{pmatrix}
 \left(\sin\theta\frac{\partial}{\partial\theta}\right)P_l(\cos\theta),
\ee
for a mode with odd parity, and 
\bea
&&h^{even}_{\mu\nu}  \nonumber\\ &=&
\begin{pmatrix}
H_0(t,r)F(r) & H_1(t,r) & 0 & 0 \\
H_1(t,r) & H_2(t,r)/F(r) & 0 & 0 \\
0 & 0 & r^2K(t,r) & 0      \\
0 & 0 & 0 & r^2K(t,r)\sin^2\theta
\end{pmatrix} \nonumber\\
&\times&  P_l(\cos\theta),
\eea
for a mode with even parity.

Here $P_l(\cos\theta)$ is the Legendre polynomial. To simplify the discussions, we will concentrate on odd parity modes only from now on. Define
\be
\psi(t,r) \equiv F(r)h_1(t,r)/r,
\ee
the equations of motion of a mode reduce to 
\be\label{eqn:eom}
\frac{\partial^2\psi}{\partial r_*^2}-\frac{\partial^2\psi}{\partial t^2}-V(r)\psi=0.
\ee
This is known as the Regge-Wheeler equation \cite{Regge:1957td}.

The black hole background is also invariant under time translation. Therefore, each frequency component of $\psi$ varies independently. Write each frequency mode as $\tilde\psi(\omega,r)$, it then satisfies the time-independent Schrodinger's equation,
\be\label{eqn:schrodinger}
\frac{d^2\tilde\psi}{d^2r_*}+(\omega^2-V(r))\tilde\psi=0,
\ee
where $r_*$ is the tortoise coordinate defined by $dr_*\equiv dr/f(r)$ and $\omega$ is the angular frequency of the mode. $V(r)$ serves as an effective gravitational potential in the form
\be\label{V_metric}
V(r)=f(r)\left(\frac{l(l+1)}{r^2}-\frac{6M}{r^3}\right),
\ee
where $l$ is the angular momentum quantum number.

\section{Coupling to the background fields}\label{sec:coup}

In general relativity, the Stuckelberg fields are purely gauge degrees of freedom 
and the perturbations should only couple to the background metric. However, 
in massive gravity theory, a complete description of the black hole solution involves
not only the metric but also the Stuckelberg fields. The gravitational perturbations 
should couple to both of them. While we lack the machinery to get the general coupling directly by perturbing massive gravity theory, here we propose a possible form of this coupling which manifests its physical effects on gravitational wave radiation even in the known Regge-Wheeler formalism we have at hand.

In Schwarzschild coordinates, the scalar mode $\pi$ depends on $t$ \cite{Berezhiani:2011mt}, thus ruining the original  Schrodinger equation form (Eq. (\ref{eqn:schrodinger})) if it couples with the perturbation. On the other hand, as we have shown, the vector mode $A^{\mu}$ depends only on $r$. Its coupling would keep the Regge-Wheeler formalism intact while manifesting its effects on gravitational perturbations only through modifying the effective potential in the Schrodinger equation. Furthermore, we consider the following simple form of coupling consistent with symmetries in theory 
\be
\mathcal{L}_{c}= \frac12\sqrt{-g}\left(\lambda A^2 \psi^2\right).
\label{Lc}
\ee
Here $\lambda$ is a dimensionless coupling constant. 

Note that in \cite{Dong:2015qpa}, when treating the coupling between a scalar perturbation field and the background Stuckelberg fields, we chose the simplest interaction between the scalar and the vector component of the Stuckelberg fields, which is normalizable and invariant under Lorentz transformations and reflections of the scalar field.\footnote{The referee brought it to our attention that the above interaction Lagrangian is not consistent with all the symmetries of the underlying theory, i.e. diffeomorphism. It does not contain derivatives of the vector field, as would be naively expected in massive gravity. In this sense, this interaction Lagrangian here only serves as an effective interaction.} Here we considered the same interaction form, treating $\psi$ as a scalar field.

We now consider the contribution to the equation of motion of $\psi$ from the above coupling term. This new interaction Lagrangian \ref{Lc} does not depend on the derivatives of $\psi$ but only depends on $\psi$ itself. Its effect is therefore only a new term in the effective potential in Eq. (\ref{eqn:schrodinger}),
\bea
\label{V_coupling}
V_{c}(r) &=& -\frac1{\sqrt{-g}\psi}\frac{\partial L_c}{\partial \psi} \\
&=& -\lambda [{A^0(r)} f(r)]^2 \\
&=& \lambda\left(\frac{M_{pl}m_g}{\kappa_0}\right)^2 \left[f(r)\int dr \frac{\sqrt{1-f(r)}}{f(r)} \right]^2.\nonumber
\eea
The integral term has logarithmic divergences at both horizons. However, it is multiplied by the $f(r)$ term in front of it which fixes this divergence, thus enabling us to use the known techniques to compute the gravitational wave emission in this case. 

To qualify how large an effect this new potential has on emitted gravitational waves, we define $B\equiv\frac{\lambda\alpha}2\left(\frac{M_{pl}}{\kappa_0 k_h}\right)^2$.  Putting $r$ in units of $1/k_h$ and $\omega$ in units of $k_h$, the strength of the effect of the new coupling is represented by $B$ in the dimensionless form of Schrodinger equation. Here $k_h$ is the surface gravity at the black hole event horizon, i.e. $k_h\equiv\frac{f'(r_h)}2$.

\section{An analytical example of gravitational ringdown}\label{sec:ana}

For a general SdS black hole, the quasinormal modes and the gravitational wave strain can only be solved numerically. In order to see how a strain comprises various quasinormal modes, we consider in the following a special case of a near-extremal SdS black hole, where the cosmological and black hole event horizons are very close, and the perturbations are only coupled to the background metric. In this case  the quasinormal modes can be analytically solved for. Moreover, for some given form of initial perturbation, the strain as a function of time can also be written in a closed form. More general cases will be dealt with numerically in the next section.

\begin{figure}[htbp!]
    \centering
      \begin{subfigure}{0.54\textwidth}
        \includegraphics[width=\textwidth]{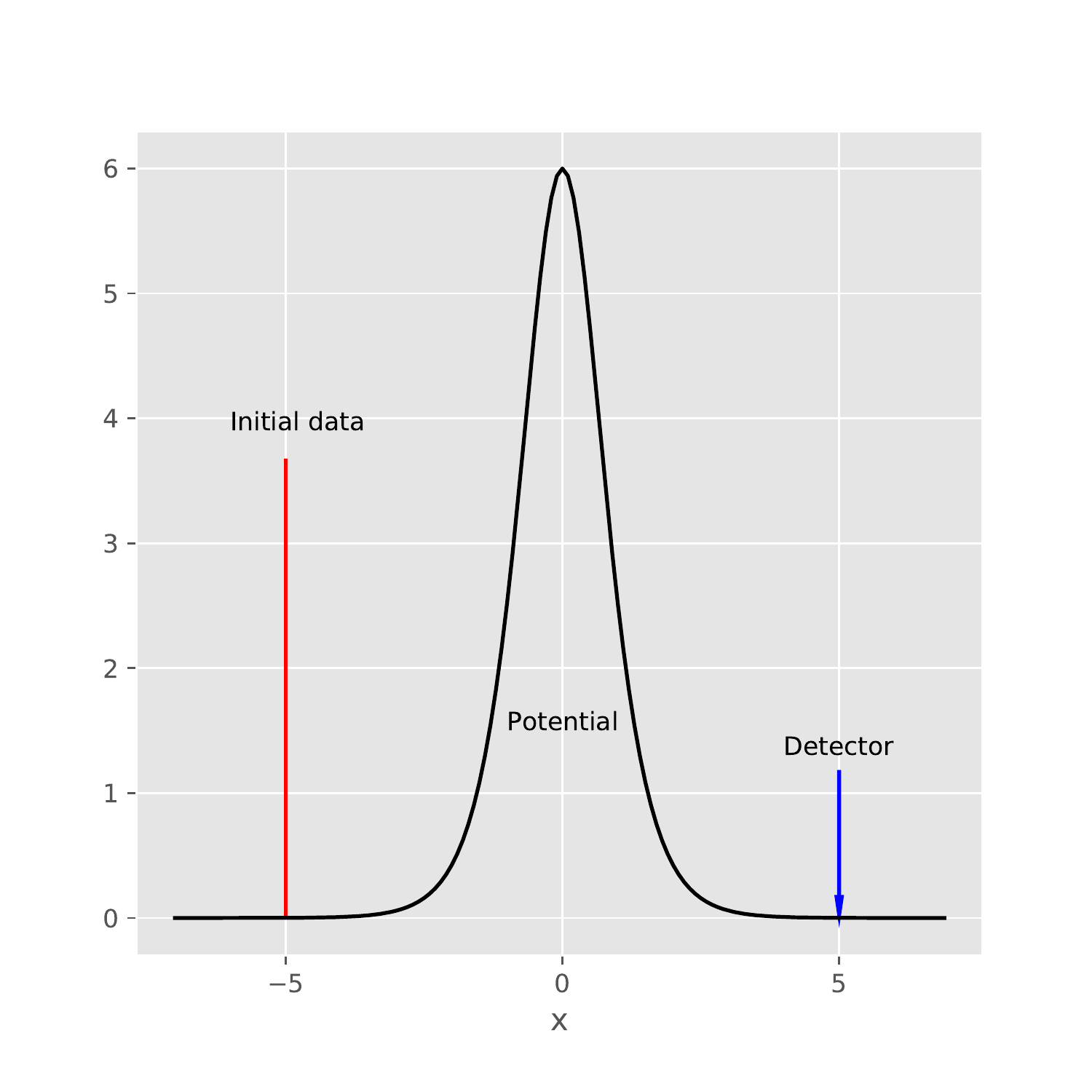}
          \caption{Potential and initial data}
          \label{fig-v-exbh}
      \end{subfigure}
      \hfill
      \begin{subfigure}{0.54\textwidth}
        \includegraphics[width=\textwidth]{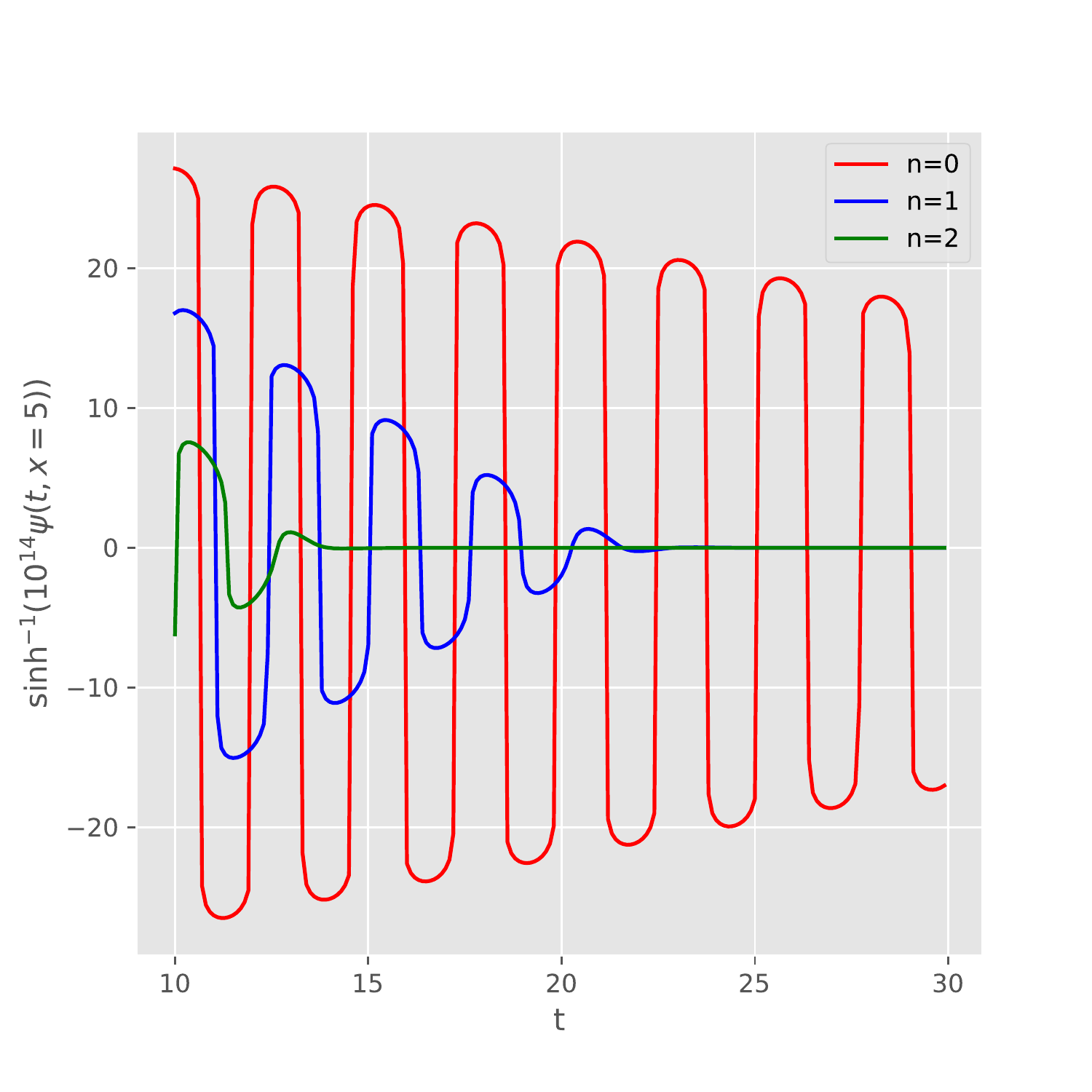}
          \caption{Strain signal}
          \label{fig-h-exbh}
      \end{subfigure}
\caption{The (\small a)  initial data and potential and (\small b) ringdown signal for a near-extremal SdS black hole. Here the tortoise coordinate x and the potential are both put dimensionless by setting $r_h=1$ and $k_h=1$. (a) The red line is the delta function initial data given by Eq. (\ref{init-delta}). The black line is the potential given in Eq. (\ref{schrodinger-ext}). The blue arrow shows the position for observing the strain signal shown below. (b) Contributions to the strain signal from individual quasinormal modes given in Eq. (\ref{qnm-ext}). Here $l=2$. $t$ starts from 10 because the initial data is put at $x=-5$ at time $t=0$ and the detector is location at $x=5$. For illustration purposes, we plot $\sinh^{-1}(10^{14}\psi)$ so that the first three quasinormal modes can be seen on the same plot \cite{Teukolsky:1974yv}.}
\label{fig-exbh}%
\end{figure}

For a near-extremal black hole, $r_{dS}\approx r_h$ and the surface gravity at $k_h$ is a small number. To order $O(k^2_h)$, the effective potential can be written as a function of $r_*$ in a closed form
\be
V(r)\approx\frac{l(l+1)}3\frac{k^2_h}{(\kappa r_h)^2\cosh^2(k_hr_*)}.
\ee
We first introduce dimensionless variables $x=k_hr_*$, $w=\omega/k_h$, and $\lambda$ which is a solution to
\be
\lambda(1+\lambda)+l(l+1)=0.
\ee
If we keep only the lowest order term in $k_h$, the Schrodinger equation is written as
\be\label{schrodinger-ext}
\frac{\partial^2\tilde\psi}{\partial x^2}+\left(w^2+\frac{\lambda(\lambda+1)}{\cosh^2x}\right)\tilde\psi=0.
\ee
The potential is plotted in Fig. (\ref{fig-v-exbh}). This equation has an analytical solution satisfying the boundary condition at the black hole event horizon, i.e. as $x\rightarrow -\infty$, as
\be
\tilde\psi_{-\infty}=[\xi(1-\xi)]^{-iw/2}  {_2}F_1(1+\lambda-iw,-\lambda-iw;1-iw;\xi),
\ee
where $\xi=(1+e^{-2x})^{-1}$, so $\xi$ goes from 0 to 1 as we go from the black hole event horizon to the cosmological horizon. $_2F_1$ is the hypergeometric function. This solution has the asymptotic behavior 
\be
\tilde\psi_{-\infty}= e^{-iw x}, x\rightarrow -\infty,
\ee
and
\be
\tilde\psi_{-\infty}= A_{in}e^{-iw x}+A_{out}e^{iw x}, x\rightarrow +\infty,
\ee
where $A_{out}$ is not important for our analysis and
\be
A_{in}=\frac{\Gamma(1-iw)\Gamma(-iw)}{\Gamma(1+\lambda-iw)\Gamma(-\lambda-iw)}.
\ee
Here $\Gamma$ is the gamma function.

Zeros of $A_{in}$ determine the quasinormal modes. $\lambda$ can be written as $-\frac12+i\lambda_i$ with $\lambda_i>0$. The quasinormal modes are then
\be\label{qnm-ext}
w_{QNM}(n)=\lambda_i-(n+\frac12)i,
\ee
for $n=0,1,2, ...$. 

To get an analytic solution for $\psi(t,x)$ near $+\infty$, we choose initial data near the black hole horizon
\be\label{init-delta}
\psi_0(x)=\delta(x-x_{-\infty}),
\ee
where $x_{-\infty}$ is a close to $-\infty$. In practice, we choose $x_{-\infty}=-5$ where the potential is effectively zero, as illustrated in Fig. (\ref{fig-v-exbh}).

The strain solution can be obtained using Eq. (\ref{strain-qnm})
\bea\label{h-t_exbh}
\psi(t,x) &=& -\frac12\sum_{n=0} (-1)^n n! \nonumber\\
&\times&\left[\frac{\Gamma(-n+2i\lambda_i) e^{[-(n+1/2)+i\lambda_i][t-(x-x_{-\infty})]}}{[-(n+1/2)+i\lambda_i]\Gamma^2\left(-(n+1/2)+i\lambda_i\right)} + c.c.\right].\nonumber\\
\eea

The result together with contributions from individual $n$'s is shown in Fig. (\ref{fig-h-exbh}), for $l=2$. We put a delta-function source at $x=-5$ at time $t=0$, and receive the strain signal at $x=5$ from $t=10$ on, as shown in the figure.

The dominant quasinormal modes decay exponentially while oscillating, with a decay rate at the same order of magnitude as the oscillation frequency. In order to show this ringing, we plotted $\sinh^{-1}(10^{14}\psi)$  as a function of $t$ while keeping $x$ fixed near the cosmological horizon (x=5). We can see that the waveform is dominated by the first quasinormal mode, while the other modes only make a subdominant contribution and are more difficult to detect in observations.

\section{Numerical calculation of gravitational waves}\label{sec:num}

In this section we perform numerical calculations for a general black hole. The perturbation is coupled both to the background metric and Stuckelberg fields, adding the new term (Eq. (\ref{V_coupling})) to the potential (Eq.(\ref{V_metric})), i.e.
\be\label{eq-v-rd10}
V\rightarrow V+V_c.
\ee

Without loss of generality, we can set $r_h=1$ and $r_{dS}=10$. The units for the potential, length, time and frequency are thus fixed. We are interested in the strain as a function of time near the cosmological horizon, given some initial perturbation near the black hole event horizon. 

\begin{figure}[htbp!]
    \centering
      \begin{subfigure}{0.54\textwidth}
        \includegraphics[width=\textwidth]{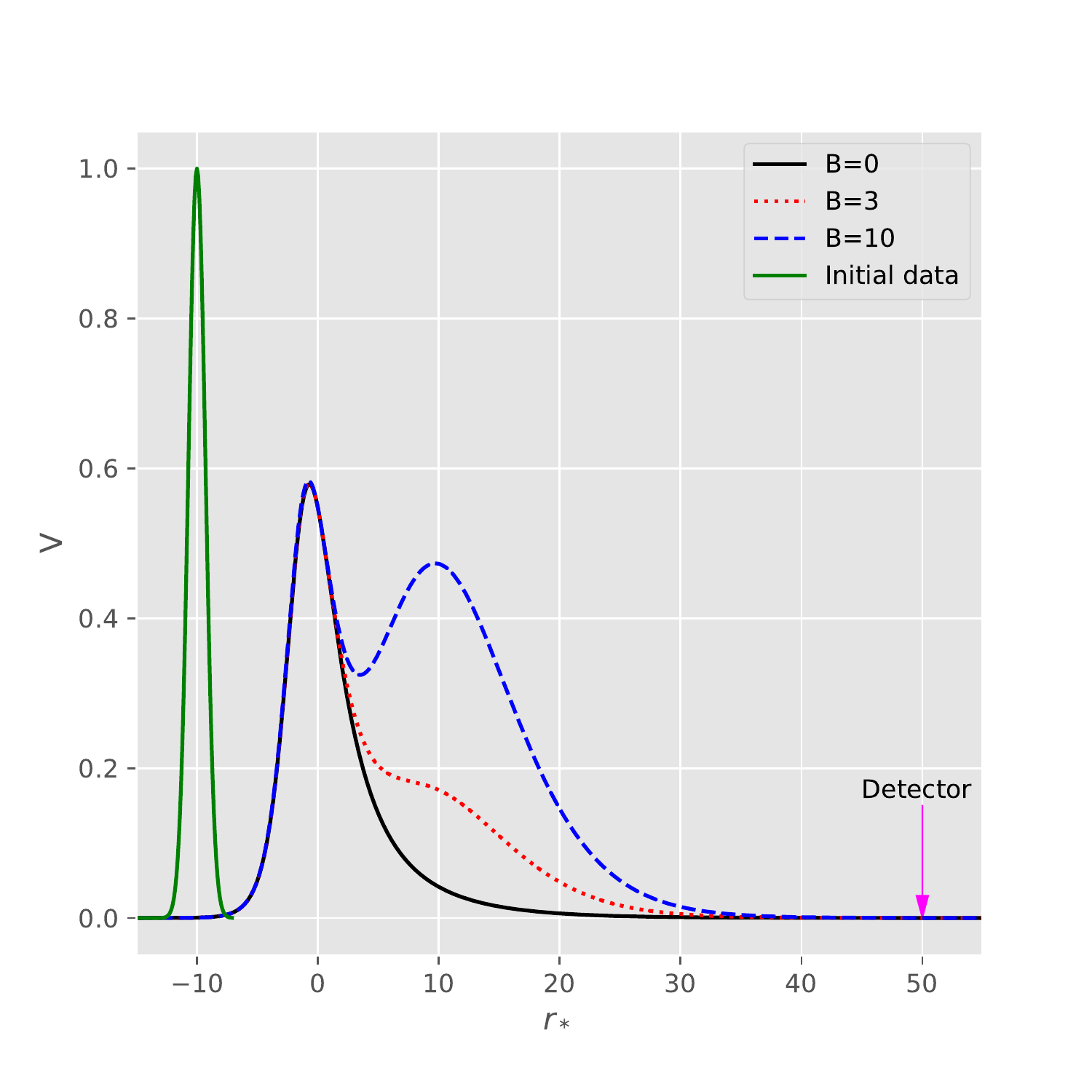}
          \caption{Potential and initial data}
          \label{fig-v-rd10}
      \end{subfigure}
      \hfill
      \begin{subfigure}{0.54\textwidth}
        \includegraphics[width=\textwidth]{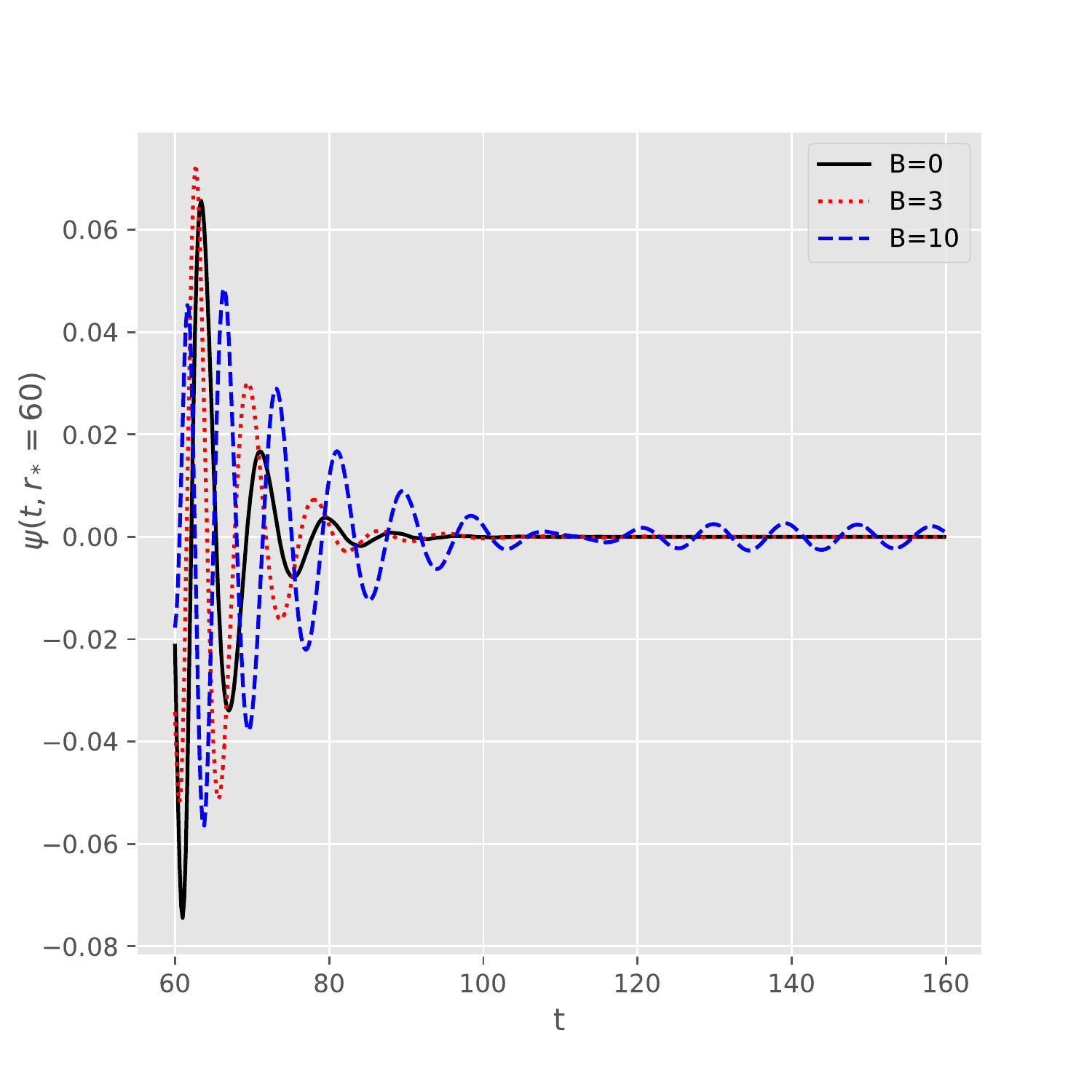}
          \caption{Strain signal}
          \label{fig-h-rd10}
      \end{subfigure}
\caption{The (\small a)  initial data and potential and (\small b) strain signal for a SdS black hole with $r_{dS}=10r_h$. Here the units are chosen by setting $r_h=1$. (a) The green solid line is the Gaussian initial data given by Eq. (\ref{init-gaussian}). The black solid, red dotted and blue dashed lines are the potential given in Eq. (\ref{eq-v-rd10}), for $B=0$, 3 and 10, respectively. In the potential, the left peak comes from the gravitational potential, and the right peak/bump comes from coupling with Stuckelberg fields. The magenta arrow shows the position for observing the strain signal shown below. (b) Expected strain signal for $B=0$, 3 and 10. Here $l=2$. The initial perturbation is around $r_*=-10$ at $t=0$ and the detector is at $r_*=50$. Therefore, our signal starts from $t=60$.}
\label{fig-exbh}%
\end{figure}

For this black hole, the effective gravitational potential is shown in Fig. (\ref{fig-v-rd10}) for $B=0$, 3 and 10, for $l=2$. We see that coupling with Stuckelberg fields effectively adds a crest to the right of the original crest in the effective potential. The height of the new crest is determined by the coupling parameter $B$. When $B$ is very small, the effect of this coupling is negligible. When $B$ is very large, the effect of the centrifugal potential is buried under the effect of the new coupling and the ringdown signal becomes totally different from the case without this new coupling. We are interested, however, in the case with intermediate values of $B$ where the similar quasinormal ringing still exists but new phenomena could occur. $B=10$ belongs to this category.

We also plotted Gaussian initial data in Fig. (\ref{fig-v-rd10}), given by 
\be\label{init-gaussian}
\psi_0(r_*)=\exp(-(r_*-r_{*0})^2),
\ee
where $r_{*0}=10$ is roughly the position of initial perturbation. The wave partly climbs over the left barrier and excites the black hole quasinormal modes. The quasinormal modes are partly received by a detector at infinity (in practice $r_*=50$). When B is not too small, the excitation is also partly reflected and transmitted at the two barriers. As a result, we also expect to observe echoes. In Fig. (\ref{fig-h-rd10}), we show the signal in the time domain for $B=0$, 3 and 10.

When the gravitational perturbation is only coupled with background metric, the potential filters out the dominant quasinormal mode from the initial data. When the coupling with background Stuckelberg fields is large enough, the effective potential shows another peak. This change in the potential results in two changes in the observed gravitational wave signal. First, the quasinormal ringing mode moves to higher oscillation frequency with lower decay rate. Second, an echo shows up at a later time. This new signal decays very slowly with time.

\begin{figure}[h]
  \centering
\includegraphics[height=0.54\textwidth,angle=0]{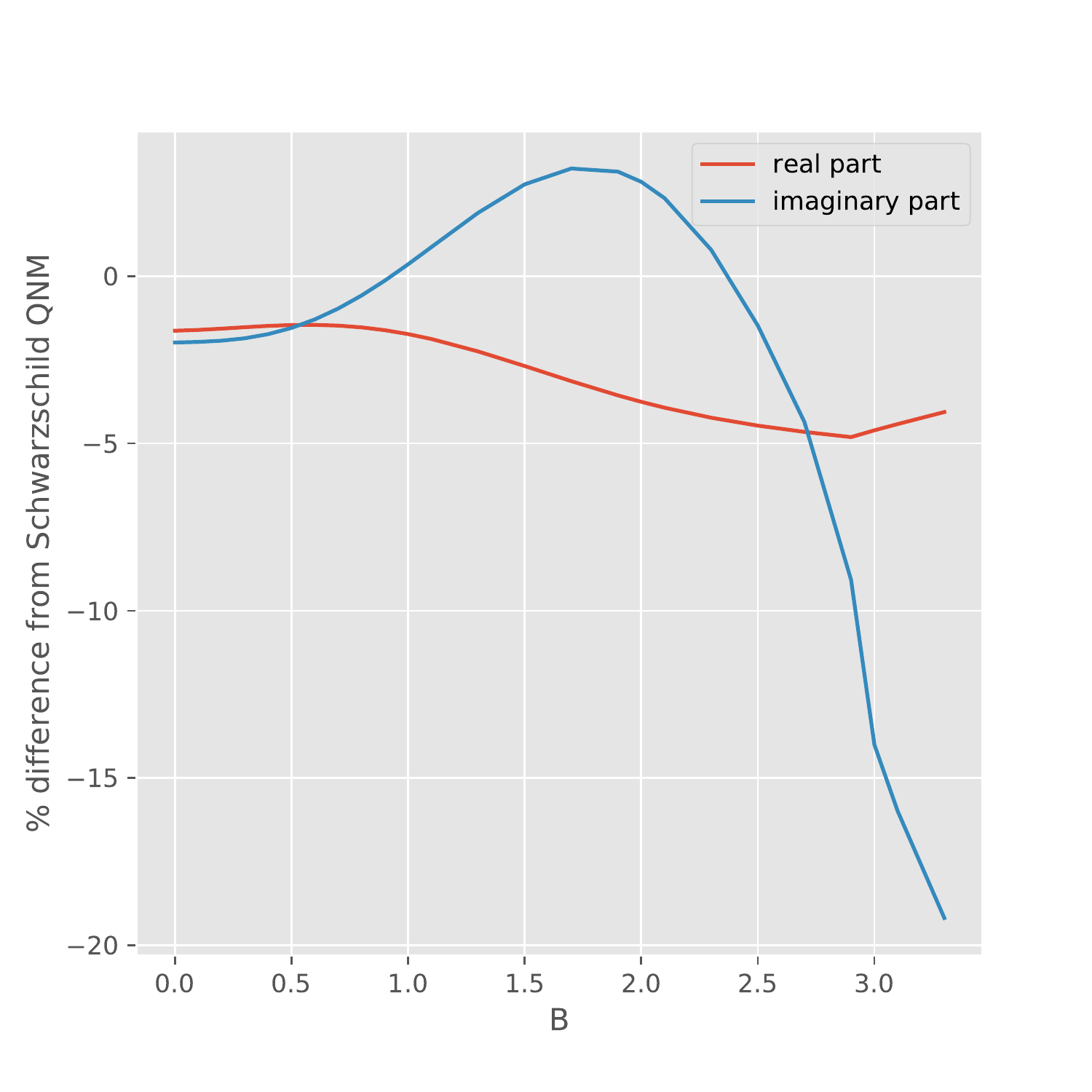}
\caption{The percent difference of the frequency of quasinormal mode (QNM) from that of a Schwarzschild black hole, as a function of the coupling strength between gravitational perturbation and background Stuckelberg fields. Here $r_{dS}=10r_h$. What is shown is  the $n=0$, $l=2$ mode. The magnetic quantum number $m$ does not enter the 
equation of motion (Eq. (\ref{eqn:eom})) and therefore can be any integer from -2 to 2.}
\label{fig-f-rd10}
\end{figure}

Observationally, the frequency of the principal quasinormal mode has drastically changed compared to a Schwarzschild black hole when B is large enough to produce echoes. But the relative amplitude of echo compared to the quasinormal mode depends on the form of initial data. For certain initial perturbations, it is still possible that the echo could be larger and thus more observable than the quasinormal mode. On the other hand, the frequency of the quasinormal mode does not depend on initial data. Deviation of it from the quasinormal mode of a Schwarzschild black hole can also serve as a test of massive gravity. The percent relative deviation is shown in Fig. (\ref{fig-f-rd10}).

We found the frequency of the principal quasinormal mode by finding a subset of the strain time series which can be best fitted with an exponentially damped sinusoid. When $B=0$, i.e. for a pure SdS black hole with $r_{dS}=10r_h$, the principal quasinormal mode is different from a Schwarzschild black hole by only 2\%. In general, the imaginary part of the mode varies more significantly with B than the real part. When B reaches 3, the imaginary part changes by 15\% while the real part changes by about 5\%.

At the end we note that an observer has to be properly placed in order to see the observational signature we describe here. dS horizon in the metric  (\ref{ds2}) is determined by the graviton mass, and it is not equal to the cosmological horizon. The constraint from GW170104 on the gravitons mass is $m_g<7.7\times 10^{-23}$ eV \cite{Abbott:2017vtc}, which translates into dS horizon greater than $1.7$ light years. On the other hand, the observed cosmological horizon size is about $10^{10}$ light years. So if we place an observer well within the graviton's dS horizon, he will observe only departure from the  Schwarzschild form of the quasinormal mode, as described in Fig.~(\ref{fig-f-rd10}). To see the echoes, he must be able to see  both barriers from the massive gravity in the potential, and therefore must be placed outside of the graviton's dS horizon.  Since most of the observed gravitational waves signals are likely to come from many millions of light years away, a generic observer would be placed outside of the graviton's dS horizon but within the cosmological dS horizon, and would be able to observe the complete signature we described here.

\section{Conclusion}\label{sec:con}
In this paper, we studied gravitational perturbations of a black hole solution in ghost-free massive gravity theory. While the black hole solution has the SdS form, there exist non-trivial Stuckelberg fields in the background. We computed the gravitational wave signal that can be observed in the asymptotic region. For perturbations coupled only with the metric, the ringdown is dominated by the first quasinormal mode. We showed this analytically for the special case where the black hole is near extremal and the source of perturbation has a delta-function form. However, this is no longer true when coupling with Stuckelberg fields is turned on. The coupling results in a separate peak/bump on top of the one caused by gravity in the effective potential. The existence of two peaks then modifies the quasinormal modes in a non-trivial way. It produces echoes in the gravitational wave time series at a later time than the quasinormal mode. In addition, it modifies the frequency of the quasinormal mode. These two signatures provide us with a clear cut  way to test massive gravity using gravitational-wave observations.

\begin{acknowledgments}
We would like to thank Vitor Cardoso for very helpful discussions and providing us the code for calculating gravitational waves in the time domain. D.S. is partially supported by the US National Science Foundation, under Grants No. PHY-1820738 and PHY-2014021.

\end{acknowledgments}

\appendix

\section{Time-domain solution of gravitaitonal ringdown}
We start with the general form of a sourceless perturbation equation in Schwarzschild or SdS background,
\be
\frac{\partial^2\psi}{\partial r_*^2}-\frac{\partial^2\psi}{\partial t^2}-V(r)\psi=0,
\ee
with initial data $\psi_0(r_*)\equiv \psi(0,r_*)$ and $\frac{\partial\psi(t,r_*)}{\partial t}\big\rvert_{t=0}=0$.

With Laplace transformation,
\be
\tilde\psi(\omega,r_*)=\int_0^\infty dt\psi(t,r_*)e^{i\omega t},
\ee
we get an ordinary differential equation with a source, i.e.
\be
\frac{\partial^2\tilde\psi}{\partial r_*^2}+(\omega^2-V(r))\tilde\psi=i\omega\psi_0(r_*).
\ee

This equation can be solved using the Green's method. The solution to
\be
\frac{\partial^2 G(\omega,r_*-r_*')}{\partial r_*^2}+(\omega^2-V(r)) G(\omega,r_*-r_*')=\delta(r_*-r_*')
\ee
is
\be\label{green}
    G(\omega,r_*-r_*')=\left\{
                \begin{array}{ll}
                  \tilde\psi_{-\infty}(\omega,r_*)\tilde\psi_{+\infty}(\omega,r_*')/W, r_*<r_*'\\\\
                  \tilde\psi_{+\infty}(\omega,r_*)\tilde\psi_{-\infty}(\omega,r_*')/W, r_*>r_*'
                \end{array}
              \right.
\ee
Here $\tilde\psi_{-\infty}$ and $\tilde\psi_{+\infty}$ are solutions to the homogeneous equation 
\be\label{homo-eqn}
\frac{\partial^2 \tilde\psi_{\pm\infty}}{\partial r^2_*} + (\omega^2-V(r))\tilde\psi_{\pm\infty} = 0.
\ee
They satisfy the boundary conditions
\bea\label{boundary}
\tilde\psi_{-\infty}&\approx& e^{-i\omega r_*}, r_*\rightarrow -\infty,\nonumber\\
\tilde\psi_{+\infty}&\approx& e^{i\omega r_*}, r_*\rightarrow +\infty,
\eea
and $W$ is the Wronskian dependent on $\omega$ only, defined as
\be
W=\det
\begin{bmatrix}
    \tilde\psi_{-\infty}       & \tilde\psi_{+\infty} \\
    \frac{\partial\tilde\psi_{-\infty}}{\partial r_*}  & \frac{\partial\tilde\psi_{+\infty}}{\partial r_*}
\end{bmatrix}.
\ee

We are interested in the strain signal at $r_*\approx +\infty$. On the other hand, the perturbation starts not far from the black hole horizon, e.g. during a binary merger. Therefore,
\bea
\tilde\psi(\omega,r_*)&=&\int dr_*'G(r_*-r_*')i\omega\psi_0(r_*')\nonumber\\
&=&\frac{i\omega\tilde\psi_{+\infty}(r_*)}{W}\int_{-\infty}^{+\infty} dr_*' \tilde\psi_{-\infty}(r_*')\psi_0(r_*').\label{psi-w-infinity}
\eea
Here we have used the property that the Wronskian is independent of $r_*$ or $r_*'$. If $\psi_{-\infty}$ has the asymptotic form
\be
\tilde\psi_{-\infty}\rightarrow A_{in}e^{-i\omega r_*}+A_{out}e^{i\omega r_*}, r_*\rightarrow +\infty,
\ee
then $W$ can be evaluated at $+\infty$ to be
\be
W=2i\omega A_{in}(\omega).\label{W}
\ee

The integral in Eq. (\ref{psi-w-infinity}) can be simplified if we assume that the source of perturbation is near the black hole horizon. Therefore, $\psi_0$ and $\tilde\psi_{-\infty}$ have common support only around $-\infty$, where $\tilde\psi_{-\infty}$ has the given form in Eq. (\ref{boundary}). The integral can be then simplified as 
\be\label{psi0-w}
\int_{-\infty}^{+\infty} dr_*' \tilde\psi_{-\infty}(r_*')\psi_0(r_*') \approx \int_{-\infty}^{+\infty} dr_*' e^{-i\omega r'_*} \psi_0(r_*') =\bar\psi_0(\omega),
\ee
where we write the fourier transform of initial perturbation as $\bar\psi_0(\omega)$.

We now perform inverse Laplace transformation, using Eqs. (\ref{psi-w-infinity}, \ref{W}, \ref{psi0-w}), to get
\be
\psi(t,r_*)=\int_{-\infty}^{+\infty}\frac{d\omega}{2\pi} \frac{\bar\psi_0(\omega) e^{-i\omega(t-r_*)}}{2A_{in}(\omega)}.\label{psi-t-infinity}
\ee

Two methods can be used to do the integration in Eq. (\ref{psi-t-infinity}). Method 1 entails calculation of quasinormal modes, which are defined as the complex simple poles of the Green's function. From Eqs. (\ref{green}, \ref{W}), they are the simple zeros of $A_{in}(\omega)$, written as $\omega^{(n)}_{QNM}$. Using contour integration, we can get 
\be\label{strain-qnm}
\psi(t,r_*)=\sum_n \Im\left( \frac{\bar\psi_0(\omega)e^{-i\omega(t-r_*)}}{A'_{in}(\omega)}\Bigg\rvert_{\omega=\omega^{(n)}_{QNM}}\right).
\ee
We used this method to obtain the analytical solution for near-extremal black hole in section \ref{sec:ana}. 

Method 2 involves computing $A_{in}(\omega)$ for real values of $\omega$ by integrating Eq. (\ref{homo-eqn}) for $\tilde\psi_{-\infty}$. Then the strain time series is obtained by a direct integration in Eq. (\ref{psi-t-infinity}). We used this method to obtain the numerical solution for a SdS black hole with $r_{dS}=10 r_h$ in section \ref{sec:num}.

\end{document}